\documentclass{ws-procs9x6}

\newcommand{\bear}{\begin{eqnarray}}

\newcommand{\beq}{\begin{equation}}
\newcommand{\eeq}{\end{equation}}

\newcommand{\la}[1]{\label{#1}}
\newcommand{\ear}{\end{eqnarray}}

\newcommand{\at}{\overline{10}}

  \def\beqr{\begin{eqnarray}}
  \def\eeqr{\end{eqnarray}}

  \def\Tr{\mbox{Tr}}

\begin{document}

\title{NOTES ON EXOTIC ANTI-DECUPLET OF BARYONS}

\author{M.V. POLYAKOV}

\address{Institut de Physique,
Universit\'e de Li\`ege, B-4000 Li\`ege 1, Belgium }

\address{Petersburg Nuclear Physics Institute, Gatchina,
St. Petersburg 188 300, Russia }

\maketitle

\abstracts{
We emphasize the importance of identifying non-exotic $SU_{\rm fl}(3)$ partners of the $\Theta^+$ pentaquark,
and indicate possible ways how to do it. We also use the soliton picture of baryons to relate Reggeon couplings
of various baryons. These relations are used to estimate the $\Theta^+$ production cross section in high
energy processes. We show that the corresponding cross sections are significantly suppressed relative
to the production cross sections of usual baryons. Finally, we present spin non-flip form factors of the
anti-decuplet baryons in the framework of the chiral quark soliton model.
}

\section{Introduction}

The first independent evidence for the exotic baryon $\Theta^+$ with strangeness $+1$ in
$\gamma\ ^{12}{\rm C}$~\cite{Osaka} and $K^+{\rm Xe}$~\cite{ITEP} reactions,
followed by important confirmation in about ten experiments by spring
2004~\cite{JLab,ELSA,neutrino,HERMES,Protvino,Juelich,Dubna,ZEUS}, urge us to take a fresh look at baryon
spectroscopy. We still know rather little about the properties of the exotic $\Theta^+$, even its very existence
is not yet firmly established, see the null experimental results~\cite{negative}. References to
other unpublished no-sighting results can be found in Refs.~\cite{LK,Poch}. In Refs.\cite{LK,AS} it was suggested that
the contradiction between {\it pro} and {\it contra} experiments is due to a particular production mechanism
of the $\Theta^+$ through a decay of the cryptoexotic $N^*(2400)$ resonance. Additionally, it was demonstrated
in Ref.\cite{AS} that the limit put by the BES collaboration on the $\Theta^+$ production in $J/\psi$ decays
is considerably higher than one may expect.

In this contribution I am neither going to review the ideas which
lead us to the prediction of the $\Theta^+$~\cite{DPP1997} nor
account for various theoretical ideas about the possible nature of
the exotic pentaquarks. For the former I can recommend the reader
the talk by D.~Diakonov at the APS meeting~\cite{diakonov}. For a
review of other theoretical ideas see the contribution by
K.~Maltman to this conference~\cite{Maltman}. Here I intend just
to stress the importance to search for non-exotic (cryptoexotic)
flavour partners of the $\Theta^+$ pentaquark. As an original
contribution, I have decided to include my notes written in 1997.
These notes appeared as the result of the discussion with
J.~Bjorken and J.~Napolitano of the possibility to search for the
$\Theta^+$ in the LASS data~\cite{Napolitano}. Probably today
these calculations can be useful to explain why the production of
the $\Theta^+$ is suppressed in some of the high energy
experiments. Also I include my old notes on the vector and scalar
form factors of the anti-decuplet baryons, which can be helpful to
understand better the anti-decuplet baryons as they emerge in the
chiral quark soliton model.

A very important point is that the discovery of a baryon with positive strangeness would
imply the existence of a new flavour multiplet of baryons, beyond the familiar octets and decuplets.
The exotic baryon has always to be accompanied by its siblings.
The minimal $SU_{\rm fl}(3)$ multiplet containing pentaquarks is the anti-decuplet of baryons.
A multiplet containing pentaquarks should also contain baryons with non-exotic ``3-quark" quantum
numbers. In the case of the anti-decuplet these are: the isodoublet of non-strange ``nucleons" and
the isotriplet of $S=-1$ $\Sigma$'s. Are they found among already known baryons or should we look for new
states? How to reveal their hidden exoticness? In our view it is very important to identify the non-exotic
partners of the $\Theta^+$ pentaquark in order to understand its nature.

To do this one can employ 1) symmetry rules dictated by the
flavour $SU(3)$, 2) the dynamical picture of the anti-decuplet
baryons. Surprisingly, this topic has not been discussed
sufficiently in the literature, although it is as important as the
pentaquark itself. Certain studies have been undertaken in {\it
e.g.} Refs.~\cite{RP,JW,DP3,AAPSW,Cohen,Glozman}. One of the
striking properties of the nucleons from the anti-decuplet is that
they can be excited by an electromagnetic probe much stronger from
the neutron target than from the proton one~\cite{RP}. Sensational
evidence for the nucleon resonance with such properties and in the
expected mass range~\cite{DP3,AAPSW} has been reported at this
conference by V.~Kouznetsov~\cite{Slava}. Further evidence for
this state has been reported by the STAR
collaboration~\cite{Kabana}. It could be that for many years we
have been overlooking a {\it narrow} nucleon resonance with the
mass around 1700~MeV! This could be possible due to the unusual
properties of this resonance inherited from its anti-decuplet
origin. It is expected~\cite{DPP1997,AAPSW} that the nucleon from
the anti-decuplet has a rather small coupling to the $\pi N$
channel, with the preferred decay channels such as $\pi\pi N, \eta
N$ and $\bar K \Lambda$. The existence of such nucleon resonance
can be clarified relatively easily with such machines as CEBAF,
MAMI, ELSA, etc. As to the $\Sigma$'s from the anti-decuplet, they
are also expected to be relatively narrow, as it follows from the
$SU_{\rm fl}(3)$ rules. Such states can be searched for in high
energy collisions, although the corresponding production cross
section can be rather strongly suppressed, see the next section.

\section{Reggeon couplings from the chiral soliton picture}

Here we derive the relations between Reggeon couplings to various baryons,
including the exotic pentaquark $\Theta^+$. Such kind of relations are
useful for estimates of the production cross sections of baryons in
high energy processes. We apply these relations to estimate
the $\Theta^+$ production cross section in the reaction
$K^+p\to \pi^+_{fast} \Theta^+\to \pi^+_{fast} K^+ n$ at $p_{\rm lab}=11.5$~Gev/c.
The corresponding data were collected by the LASS collaboration, see {\it e.g.}
Ref.~\cite{Napolitano}.

We restrict ourselves to the spin-flip-dominated production reactions:
\begin{itemize}
\item
$\pi^-p\to \pi^0 n $,
\item
$\pi^+ p \to \pi^0\Delta^{++}$,
\item
$\pi^+p\to K^+\Sigma^{*+}(1385)$,
\item
$K^- p\to
\pi^-\Sigma^{*+}(1385)$,
\item
$K^+p\to \pi^+ \Theta^+$ ,
\end{itemize}
Other spin-flip-dominated reactions can be related to these by the
(broken) $SU(3)$ relations for the Reggeon couplings, which are known
to work well (see Ref.~\cite{Irving}). In the chiral quark soliton model the
low-lying baryons are different rotational excitations of the
same object.  This enables us to derive relations
between spin-flip Reggeon couplings in the above list of reactions.
 We shall check the relations between Reggeon coupling from the chiral
 soliton confronting them with the data on measured reactions from the above
 list.  These relations can be used to estimate the
 production cross section of the exotic $\Theta^+$ baryon, say, in the reaction
 $K^+p\to \pi^+_{fast} \Theta^+\to \pi^+_{fast} K^+ n$~\cite{Napolitano}.

We consider here only the spin-flip dominated reactions, since
our objective is to estimate the production cross section of the
 exotic $\Theta^+$ baryon in the reaction $K^+p\to \pi^+ \Theta^+$ which is obviously
 spin-flip dominated (the spin non-flip part is zero for transitions between
 baryons from different $SU(3)$ multiplets, this was confirmed by
 experiment: the spin non-flip part of the amplitude of, say,
 $\pi^+p\to K^+\Sigma^{*+}(1385)$ and $\pi^+ p \to \pi^0\Delta^{++}$ reactions is
 negligibly small). The smallness of the spin non-flip part of the
 amplitude of the reaction $\pi^-p\to \pi^0 n $ is related to the large
 isovector magnetic moment of the nucleon.

 The soliton-Reggeon coupling can
 be written in terms of the rotational coordinates $R$ of the baryon as
 (for notations see Ref.~\cite{DPP1997})

\beq  3 w_0\ \frac 12
\Tr(R^\dagger\lambda^m R\lambda_3)\sqrt{-\alpha' t}\cdot
\frac{1-\mbox{e}^{-i\pi\alpha{(t)}}}{\sin{\pi \alpha(t)}}\, .
\label{meson-soliton}
\eeq
Here $\alpha(t)$ is the corresponding Regge trajectory and index $m$ denotes
the flavour of the leading meson on the corresponding trajectory ($\rho, K^*$).

In the next-to-leading order we have to add to
eq.~(\ref{meson-soliton}) collective operators depending on the
angular momentum $J_a$. The corresponding operators have the form :

\bear
\nonumber
&&
\Biggl[ -i 3 w_1 \cdot \frac 12
d_{3\alpha\beta} \Tr(R^\dagger\lambda^m R\lambda_\alpha) J_\beta
-
 \frac{-i 3 w_2}{\sqrt 3}\cdot \frac 12
\Tr(R^\dagger\lambda^m R\lambda_8) J_3 \Biggl]\\
 &\times&\sqrt{-\alpha' t}\cdot
\frac{1-\mbox{e}^{-i\pi\alpha{(t)}}}{\sin{\pi \alpha(t)}},
\la{meson-soliton-rot}\ear
where $d_{abc}$ is the $SU(3)$ symmetric tensor, $\alpha, \beta=4,5,6,7$ and $J_a$ are the generators
of the infinitesimal $SU(3)$ rotations.

Sandwiching eqs.~(\ref{meson-soliton},\ref{meson-soliton-rot})
between the rotational wave functions of the initial and final baryons
(the explicit expressions for the corresponding wave functions can be
found in Appendix~A of Ref.~\cite{DPP1997}), one gets
the following expressions for the $B_1\rightarrow B_2+{\rm Reggeon}$ vertices in the reactions
listed above (we omit the kinematical factors):

\framebox{$\rho^-pn$} \beq -i 3 G_8 \frac{7\sqrt
2}{30}\ ,
\label{ver1} \eeq \framebox{$\rho^+p\Delta^{++}$} \beq -i 3
G_{10} \frac{1}{\sqrt 5}\  C_{\frac 12 S_3;1 0}^{\frac 32 S_3}\, ,
\label{ver2} \eeq \framebox{$\bar K^{*0}p\Sigma^{*+}$} \beq -i 3
G_{10}\frac{1}{\sqrt{15}}\
  C_{\frac 12 S_3;1 0}^{\frac 32
S_3}\, , \label{ver3} \eeq \framebox{$K^{*0}p\Theta^+$}
\beq -i 3 G_{\at} \frac{1}{\sqrt{30}}\, .
\label{ver4} \eeq
Here we introduced the following coupling constants:
\bear \nonumber G_8&=&w_0-\frac 12 w_1-\frac{1}{14}
w_2 \, ,\\ \nonumber G_{10}&=&w_0-\frac 12 w_1 \, ,  \\ \nonumber
G_{\at}&=&w_0+ w_1+\frac{1}{2} w_2 \, . \ear

The constants $w_{i}$ can be estimated using the measured high energy processes.
We shall be interested in the ratios of various cross sections, therefore for
us here only the ratios of these constants are relevant.
The structure of the collective operators (\ref{meson-soliton},\ref{meson-soliton-rot})
is the same as in the case of the axial and vector currents. The analysis of the corresponding
axial and magnetic constants~\cite{Kim:axial,Kim:magnetic} indicates that the ratios $w_{1,2}/w_0$
are negative. Model calculations~\cite{Kim:magnetic} confirm the negative sign of
$w_{1,2}/w_0$ and give the following values:

\beq \frac{w_1}{w_0}= -0.35 \pm 0.1 \qquad \frac{w_2}{w_0}= -0.25
\pm 0.1 \; ,
 \label{numest}
\eeq
where the errors are added {\em by hands} simply on the basis of
our working experience with this model. It should be mentioned that the
non-relativistic quark model (which, to some extent, can be used
as a guiding line) predicts $w_1/w_0=-4/5$ and $w_2/w_0=-2/5$,
which is in a qualitative agreement with a more realistic
calculation in the quark soliton model. Amazingly, though, these
ratios produce exactly zero $G_{\at}$. At the moment we are unable
to point out the deep reason for such cancellation.

Using the equations derived above we can obtain the
relations between the cross
sections of different spin-flip dominated reaction
 (the list is given at the beginning of the section). In doing this, we shall assume
 that these reactions are dominated by the one Reggeon exchange ($\rho$
 and $K^*$-trajectories). The first group of relations is simply
 the $SU(3)_{\rm fl}$ relations which are known~\cite{Irving} to be well
     reproduced by the experimental data. Given this fact, we shall not
     discuss this group of relations. The nontrivial prediction of the
 chiral quark-soliton model is the relations between the high energy
     reactions which involve baryons from {\em different} $SU_{\rm fl}(3)$ multiplets.
     These are (for the same incident $p_{\rm lab}$):

     \beq
\frac{\sigma(\pi^+ p \to \pi^0\Delta^{++})}{\sigma(\pi^-p\to \pi^0
n)}= \frac{60}{49}\cdot \frac{(w_0-\frac 12 w_1)^2}{(w_0-\frac 12
     w_1-\frac{1}{14}w_2)^2} \, ,
     \label{frel}
     \eeq

     \beq
\frac{\sigma(K^+p\to \pi^+ \Theta^+)}{\sigma(\pi^+p\to
K^+\Sigma^{*+}(1385))} =\frac{\sigma(K^+p\to \pi^+
\Theta^+)}{\sigma(K^-p\to \pi^-\Sigma^{*+})}= \frac{3}{4}\cdot
\frac{(w_0+w_1+\frac 12 w_2)^2}{(w_0-\frac 12
     w_1)^2} \, .
     \label{srel}
     \eeq
All other relations can be obtained with the help of the (broken)
$SU(3)$ relations and hence they are trivial. Eq.~(\ref{frel})
     can be confronted with experiment, whereas eq.~(\ref{srel})
     is the prediction. Let us note that the first equation in (\ref{srel})
is a consequence of the assumed exchange degeneracy of Regge
trajectories. The exchange degeneracy is in general violated,
although not very strongly; for a rough estimate of the $\Theta^+$
production cross section it is sufficient to assume the exchange
degeneracy.

 Using the estimates (\ref{numest}) for $w_{1,2}$ we obtain:

     \beq
\frac{\sigma(\pi^+ p \to \pi^0\Delta^{++})}{\sigma(\pi^-p\to \pi^0
n)}
 \approx
1.2\, .
     \label{frelnum}
     \eeq
 We see that this number is not sensitive to the uncertainties in the
 determination of $w_{1,2}$.
 Let us compare this prediction with the data,
 Ref.~\cite{Stirling} gives:

 \beq
 \int_0^{-0.5}dt \frac{d\sigma(\pi^-p\to \pi^0 n)}{dt}=
 87\pm 4 \ \mu\mbox{barn}\, ,
 \eeq
at $p_{\rm lab}=5.9$~GeV/c. The $\Delta^{++}$ production experiment
 \cite{Bloodworth} gives:

 \beq
 \int_0^{-0.5}dt \frac{d\sigma(\pi^+ p \to \pi^0\Delta^{++})}{dt}=
 133 \pm 13\ \mu\mbox{barn}\, ,
 \eeq
at $p_{\rm lab}=5.45$~GeV/c. This cross section can be rescaled to
 $p_{\rm lab}=5.9$~GeV/c using $\sigma(\pi^+ p \to \pi^0\Delta^{++})\sim
 p_{\rm lab}^{-1.59}$ \cite{Bloodworth}. Eventually we get:

     \beq
\frac{\sigma(\pi^+ p \to \pi^0\Delta^{++})}{\sigma(\pi^-p\to \pi^0
n)}= 1.35\pm0.15 \quad \mbox{(expt. at $p_{\rm lab}=5.9$~GeV/c)}  \, ,
     \eeq
 in a good agreement with our prediction (\ref{frelnum}).
The agreement is even better for experiments at higher energies.
The value of $\sigma(\pi^+ p \to \pi^0\Delta^{++})=44.8\pm 7\
\mu$ barn measured at $p_{\rm lab}=13.1$~GeV/c \cite{Scharen} being
divided by $\sigma(\pi^-p\to \pi^0 n)=36\pm 2 \ \mu$barn measured
at $p_{\rm lab}=13.3$~GeV/c \cite{Stirling} gives:

     \beq
\frac{\sigma(\pi^+ p \to \pi^0\Delta^{++})}{\sigma(\pi^-p\to \pi^0
n)}= 1.2\pm 0.1 \quad \mbox{(expt. at $p_{\rm lab}\approx 13$~GeV/c)}
\, .
     \eeq
We see that the chiral soliton model successfully predicts
non-trivial relations between Reggeon couplings to baryons from
different multiplets.

 Given this success, we turn now to the estimate of the $\sigma(K^+p\to \pi^+
\Theta^+)$. From eq.~(\ref{srel}) and the estimates of $w_{1,2}$
(\ref{numest}) we get:

\beq \frac{\sigma(K^+p\to \pi^+ \Theta^+)}{\sigma(\pi^+p\to
K^+\Sigma^{*+}(1385))}= \frac{\sigma(K^+p\to \pi^+
\Theta^+)}{\sigma(K^-p\to \pi^-\Sigma^{*+})}= 0.05\div 0.25\, ,
 \label{srelnum}
 \eeq
 we see that in this case the result is very sensitive to the uncertainties
 of $w_{1,2}$ due to the deep cancellation of these constants.
With the present state of art we can not say precisely how deep is this
cancellation, but in any case we can conclude that the suppression is rather
strong. We note that the estimate of the ratio (\ref{srelnum}) on the upper side of
0.25 is really the highest number one can get, whereas on the low side the cancellation can be much
deeper.

 In order to estimate the absolute value of
 $\sigma(K^+p\to \pi^+ \Theta^+)$ at $p_{\rm lab}=11.5$~GeV/c
\cite{Napolitano} we need the value of
 $\sigma(\pi^+p\to K^+\Sigma^{*+}(1385))$ or
 $\sigma(K^-p\to \pi^-\Sigma^{*+})$ at the same $p_{\rm lab}$.
 Fortunately these cross sections were measured exactly at
 $p_{\rm lab}=11.5$~GeV/c with the results~\cite{Baker}:
 $$\sigma(\pi^+p\to K^+\Sigma^{*+}(1385))\approx 8  \ \mu\mbox{barn}
 \, ,$$
 and~\cite{Ballow}
 $$\sigma(K^-p\to \pi^-\Sigma^{*+})\approx 10.1\pm 1.1  \ \mu\mbox{barn}
 \, .$$
A slight difference in the above two cross sections is related to the
 violation of the exchange degeneracy of the $K^*$ and $K^{**}$ trajectories.
From the above data and
 with the help of eq.~(\ref{srelnum}) we get the estimate:

 \beq
 \sigma(K^+p\to \pi^+ \Theta^+)\approx 0.5\div 2.5 \ \mu\mbox{barn}\, .
 \eeq
We see that the $\Theta^+$ production cross section is rather small.

 Let us note that the above estimate of the $\Theta^+$ production cross section
 should be considered as an order of magnitude estimate (up to a factor
 of 2). In the course of derivation we have neglected:
\begin{itemize}
\item
 The violation of the exchange degeneracy of the Regge trajectories.
 This could give an uncertainty about 20-30\%.
\item
 The  mass dependence of produced particle on the Reggeon parameters,
 {\it e.g.} we take the scale parameters $s_0$, the
 parameter in the Reggeon residue, etc. to be universal ($=\alpha'$)
 following the Veneziano model pattern.
 This could give an uncertainty about 30-40\%.
\end{itemize}

Further we note that the estimates of Reggeon couplings presented here can be used for
the calculations of the $\Theta^+$ yields in inclusive high energy processes such as
$pp\to \Theta^+ X$, especially in the triple Reggeon limit.

\section{Vector and scalar form factors of the anti-decuplet baryons}

Here we consider baryon spin non-flip form factors of exotic anti-decuplet baryons.
We give relations between various form factors in the chiral quark soliton model
neglecting the $SU_{\rm fl}(3)$ breaking effect, since our aim just to get an idea
about the qualitative behaviour of these form factors. The effects of symmetry breaking
can be easily added. We introduce notations:

\beq
\langle B| \bar \psi_f \Gamma \psi|B\rangle= F_f^{(B)}(t)\ ,
\eeq
where $f$ denotes quark flavour, $\Gamma=\gamma_0$ or $\Gamma=1$ are the Dirac
matrices corresponding to the spin non-flip vector and scalar form factors, $t$ is the momentum transfer
squared~\footnote{We do not specify $\Gamma$ in the
form factors as the relations presented below are fulfilled both for vector and scalar form factors.}.
In the chiral quark soliton model one can write a universal operator in the collective coordinate
space\footnote{Its form is similar to the mass operator written in Ref.~\cite{DPP1997}}.
This operator can be parametrized in terms of three universal form factors. Therefore
we can relate $F_f^{(B)}(t)$ for baryons from various multiplets. Let us just list some of these relations:

\noindent
\underline{Octet baryons}
\bear
 F_u^{(\Lambda)}&=& F_d^{(\Lambda)}=\frac 16 \left(4 F_d^{(p)}+F_u^{(p)}+F_s^{(p)}\right)\\
 F_s^{(\Lambda)}&=&\frac 13 \left(-F_d^{(p)}+2 F_u^{(p)}+2F_s^{(p)}\right)\\
 F_u^{(\Sigma^+)}&=&F_u^{(p)},\ F_d^{(\Sigma^+)}=F_s^{(p)},\ F_s^{(\Sigma^+)}=F_d^{(p)}\\
F_u^{(\Xi^0)}&=&F_d^{(p)},\ F_d^{(\Xi^0)}=F_s^{(p)},\ F_s^{(\Xi^0)}=F_u^{(p)}
\ear

\noindent
\underline{Anti-decuplet}
\bear
F_u^{(\Theta^+)}&=& F_d^{(\Theta^+)}=\frac{1}{12} \left(8 F_d^{(p)}+8 F_u^{(p)}-7 F_s^{(p)}\right)\\
F_s^{(\Theta^+)}&=&\frac{1}{6} \left(-2 F_d^{(p)}-2 F_u^{(p)}+13 F_s^{(p)}\right)\\
F_u^{(p^*)}&=&\frac{1}{12} \left(8 F_d^{(p)}+8 F_u^{(p)}-7 F_s^{(p)}\right)\\
F_u^{(p^*)}&=&\frac{1}{3} \left(F_d^{(p)}+ F_u^{(p)}+F_s^{(p)}\right)\\
F_s^{(p^*)}&=&\frac{5}{4} F_s^{(p)}\\
F_u^{(\Xi^+)}&=&\frac{1}{12} \left(8 F_d^{(p)}+8 F_u^{(p)}-7 F_s^{(p)}\right)\\
F_u^{(\Xi^+)}&=&\frac{1}{6} \left(-2F_d^{(p)}-2F_u^{(p)}+13 F_s^{(p)}\right)\\
F_s^{(\Xi^+)}&=&\frac{5}{4} \left(8 F_d^{(p)}+8 F_u^{(p)}-7 F_s^{(p)}\right)
\ear
All other relations can be easily obtained with the help of the U-, V- and isospin symmetries.

The detailed analysis of the above relations will be presented elsewhere. Here we just note that,
if we neglect the strange form factor of the proton, the typical radius of the $\Theta^+$ is $\sqrt{r_p^{2}+r_n^{2}}$.
Here $r_{p,n}^2$ are the electromagnetic radii of the proton and the neutron. In other words, it seems that
$\Theta^+$ is a compact object.
For instance its electric form factor is given in terms of the electric form factors of the nucleons as
$G_E^{\Theta^+}(t)=G_E^p+G_E^n-\frac 14 G_E^s$~\footnote{As it follows from the U-spin symmetry
the same relations hold for the electric form factors of the $p^*$ and $\Xi^+$.}
(we remind that the breaking of the $SU_{\rm fl}(3)$ is not taken into account).
It is also very interesting that the strange quark distribution
in the anti-decuplet nucleon follows (up to the factor 5/4) the distribution of the strange quarks in
the usual nucleon, suggesting that the additional $s$ and $\bar s$ in the $p^*$ ``sit'' close to each other.


\begin{thebibliography}{99}

\bibitem{Osaka}
T. Nakano (LEPS Collaboration), Talk at the PANIC 2002 (Oct. 7, 2002, Osaka);
T. Nakano et al., Phys. Rev. Lett. {\bf 91}, 012002 (2003), [ hep-ex/0301020].

\bibitem{ITEP}
V.A. Shebanov (DIANA Collaboration), Talk at the Session of the Nuclear Physics
Division of the Russian Academy of Sciences (Dec. 3, 2002, Moscow);
V.V. Barmin, A.G. Dolgolenko et al., Phys. Atom. Nucl. {\bf 66}, 1715 (2003)
[Yad. Fiz. {\bf 66}, 1763 (2003)], [hep-ex/0304040].

\bibitem{JLab}
S. Stepanyan, K. Hicks et al. (CLAS Collaboration), Phys. Rev. Lett. {\bf 91}, 252001 (2003);
V. Kubarovsky et al. (CLAS Collaboration), Phys. Rev. Lett. {\bf 92}, 032001 (2004),
[hep-ex/0311046].

\bibitem{ELSA}
J. Barth et al. (SAPHIR Collaboration), Phys. Lett B {\bf 572}, 127 (2003), [hep-ex/0307083].

\bibitem{neutrino}
A.~E.~Asratyan, A.~G.~Dolgolenko and M.~A.~Kubantsev,
Phys.\ Atom.\ Nucl.\  {\bf 67} (2004) 682
[Yad.\ Fiz.\  {\bf 67} (2004) 704]
[arXiv:hep-ex/0309042].

\bibitem{HERMES}
A. Airapetian et al. (HERMES Collaboration), Phys. Lett. B {\bf 585}, 213 (2004),
[hep-ex/0312044].

\bibitem{Protvino}
A. Aleev et al. (SVD Collaboration), [hep-ex/0401024].

\bibitem{Juelich}
M. Abdel-Bary et al. (COSY-TOF Collaboration), [hep-ex/0403011].

\bibitem{Dubna}
P.Zh. Aslanyan, V.N. Emelyanenko, G.G. Rikhkvitzkaya, [hep-ex/0403044].

\bibitem{ZEUS}
S. Chekanov et al. (ZEUS Collaboration), [hep-ex/0403051].

\bibitem{negative}
J.~Z.~Bai {\it et al.} [BES Collaboration], hep-ex/0402012.\\
K.~T.~Knoepfle, M.~Zavertyaev, and T.~Zivko [HERA-B
Collaboration], contribution to Quark Matter 2004;
hep-ex/0403020.\\
C.~Pinkenburg [PHENIX Collaboration], contribution to the 17th
Intern. Conf. on Ultra-Relativistic Nucleus-Nucleus Collisions,
Jan.2004; nucl-ex/0404001.

\bibitem{LK}
M.~Karliner and H.~J.~Lipkin,
arXiv:hep-ph/0405002.

\bibitem{Poch}
J.~Pochodzalla,
arXiv:hep-ex/0406077.

\bibitem{AS}
Y.~I.~Azimov and I.~I.~Strakovsky,
arXiv:hep-ph/0406312.




\bibitem{DPP1997}
D. Diakonov, V. Petrov and M. Polyakov, Z. Phys. A {\bf 359}, 305 (1997),
[hep-ph/9703373].

\bibitem{diakonov}
D.~Diakonov,
arXiv:hep-ph/0406043.

\bibitem{Maltman}
K.~Maltman, contribution to this conference;\\
see also:
B.~K.~Jennings and K.~Maltman,
Phys.\ Rev.\ D {\bf 69} (2004) 094020
[arXiv:hep-ph/0308286].



   \bibitem{Napolitano}
J.~Napolitano, J.~Cummings and M.~Witkowski,
{\it ``Baryon excitation in $K^\pm p$ reactions,''}
PiN Newslett.\  {\bf 13} (1997) 276.

\bibitem{RP}
M.~V.~Polyakov and A.~Rathke,
Eur.\ Phys.\ J.\ A {\bf 18} (2003) 691
[arXiv:hep-ph/0303138].



\bibitem{JW}
R.~L.~Jaffe and F.~Wilczek,
Phys.\ Rev.\ Lett.\  {\bf 91} (2003) 232003
[arXiv:hep-ph/0307341].

\bibitem{DP3}
D.~Diakonov and V.~Petrov,
Phys.\ Rev.\ D {\bf 69} (2004) 094011
[arXiv:hep-ph/0310212].

\bibitem{AAPSW}
R.~A.~Arndt, Y.~I.~Azimov, M.~V.~Polyakov, I.~I.~Strakovsky and R.~L.~Workman,
Phys.\ Rev.\ C {\bf 69} (2004) 035208
[arXiv:nucl-th/0312126].


\bibitem{Cohen}
T.~D.~Cohen,
arXiv:hep-ph/0402056.

\bibitem{Glozman}
L.~Y.~Glozman,
arXiv:hep-ph/0309092.

\bibitem{Slava}
V.~Kouznetsov [for the GRAAL collaboration], contribution to this conference;\\
see also: V.~Kouznetsov, talk at international workshop
{\it Pentaquark states: structure and properties},
February 10 - February 12, 2004, Trento, Italy.\\
http://www.tp2.rub.de/talks/trento04/index.html

\bibitem{Kabana}
S.~Kabana  [STAR Collaboration],
arXiv:hep-ex/0406032.




    \bibitem{Irving} A.~C.~Irving and R.~P.~Worden,
Phys.\ Rept.\  {\bf 34} (1977) 117.


   \bibitem{Kim:axial}
H.~C.~Kim, M.~Praszalowicz and K.~Goeke,
Phys.\ Rev.\ D {\bf 61} (2000) 114006
[arXiv:hep-ph/9910282].
\bibitem{Kim:magnetic}
H.~C.~Kim, M.~Praszalowicz, M.~V.~Polyakov and K.~Goeke,
Phys.\ Rev.\ D {\bf 58} (1998) 114027
[arXiv:hep-ph/9801295].



   \bibitem{Stirling} Stirling, {\em et al.},
Phys. Rev. Lett.{\bf 14} (1965) 763.
\bibitem{Scharen} J.H. Scharenguivel, {\em et al.},
Nucl. Phys. {B36}(1972) 363.
\bibitem{Bloodworth} I.J. Bloodworth {\em et al.},
Nucl. Phys. {B81} (1974) 231.
   \bibitem{Baker}
   P.A. Baker, J.S. Chima, P.J. Dornan, D.J. Gibbs, G. Hall, D.B. Miller,
T.S. Virdee, A.P. White, Nucl.Phys. {\bf B166} (1980) 207;\\
J. Ballam, J. Bouchez, {\em et al.},
{\it Amplitude analysis of $Y^* (1385)$ production in the line reversed
reactions: $\pi^+ p \to  K^+ Y^* (1385)$ and $K^- p \to \pi^- Y^* (1385)$ at 7-Gev/c
and 11.5-GeV/c.},
SLAC-PUB-2175, Aug 1978. 18pp. Contributed paper to 19th Int. Conf. on
High Energy Physics, Tokyo, Japan, Aug 23-30, 1978.
\bibitem{Ballow} J. Ballow {\em et al.},
Phys. Rev. Lett.{\bf 41} (1978) 676.
\end{thebibliography}
\end{document}